%% file: main.tex
\documentclass{article}
\pdfoutput=1
\usepackage{subcaption,rotating}
\usepackage[utf8]{inputenc}
\usepackage[margin=1.5in]{geometry}
\usepackage{amsmath}\usepackage{amsfonts}\usepackage{amssymb}\usepackage{mathrsfs}\usepackage{amsthm}\usepackage{mathtools}\usepackage{natbib}
\usepackage{graphicx}
\usepackage{listings}\usepackage{color}
\usepackage{tikz}
\usepackage{float}

\newcommand{\y}{y}

\newcommand{\be}{\begin{equation}}
\newcommand{\ee}{\end{equation}}

\usepackage{color}

\renewcommand\r[1]{#1}\renewcommand\r[1]{{\color{red}#1}}
\renewcommand\b[1]{{\color{blue}#1}}

\def\thep{F}\def\thep{p}
\def\citep{\cite}

\title{{Deep Local Volatility\footnote{A Python notebook, compatible with Google Colab, and accompanying data are available in https://github.com/mChataign/DupireNN. Due to file size constraints, the notebook must be run to reproduce the figures and results in this article.}}}

\author{Marc Chataigner\footnotemark[1] \and St\'{e}phane Cr\'{e}pey\thanks{Department of Mathematics, University of Evry,  Paris Saclay; stephane.crepey@univ-evry.fr, marc.chataigner@univ-evry.fr. The PhD thesis of Marc Chataigner is co-funded
by the Research Initiative ``Mod\'elisation des march\'es actions, obligations
 et d\'eriv\'es'',
  financed by HSBC France under the aegis of the Europlace Institute of Finance, and by a public grant as part of investissement d'avenir project, reference ANR-11-LABX-0056-LLH LabEx LMH. The views and opinions expressed in this paper are those of the authors
alone and do not necessarily reflect the views or policies of HSBC Investment
 Bank, its subsidiaries or affiliates.}
\and Matthew Dixon\thanks{Department of Applied Mathematics, Illinois Institute of Technology, Chicago; matthew.dixon@iit.edu.}}

\begin{document}

\maketitle

\abstract{Deep learning for option pricing has emerged as a novel methodology for fast computations with applications in calibration and computation of Greeks. However, many of these approaches do not enforce any no-arbitrage conditions, and the subsequent local volatility surface is never considered. In this article, we develop a deep learning approach for  
 interpolation of European vanilla option prices which jointly yields the full surface of local volatilities. We demonstrate the modification of the loss function or the feed forward network architecture to enforce (hard constraints approach) or favor (soft constraints approach) the no-arbitrage conditions and we specify the experimental design parameters that are needed for adequate performance. A novel component is the use of the Dupire formula to enforce bounds on the local volatility associated with option prices, during the network fitting. Our methodology is benchmarked numerically on real datasets of DAX vanilla options.}

 
\newcommand{\dK}[1]{\frac{\partial #1}{\partial K}}\renewcommand{\dK}[1]{\partial_K #1}

\newcommand{\dT}[1]{\frac{\partial #1}{\partial T}}\renewcommand{\dT}[1]{\partial_T #1}

\newcommand{\gK}[1]{\frac{\partial^2 #1}{\partial K^2}}\renewcommand{\gK}[1]{\partial^2_{K^2} #1}

\newcommand{\dk}[1]{\frac{\partial #1}{\partial k}}\renewcommand{\dk}[1]{\partial_k #1}

\newcommand{\gk}[1]{\frac{\partial^2 #1}{\partial k^2}}\renewcommand{\gk}[1]{\partial^2_{k^2} #1}

\def\sp{\,,\;}
\renewcommand\r[1]{#1}\renewcommand\r[1]{{\color{red}#1}}
\renewcommand\b[1]{{\color{blue}#1}}
 

\section{Introduction}\label{sect:intro}
A recent approach to option pricing involves reformulating the pricing problem as a surrogate modeling problem. Once reformulated, the problem is amenable to standard supervised machine learning methods such as Neural networks and Gaussian processes. This is particularly suitable in situations
with a need for fast computations and a tolerance to approximation error.
In their seminal paper, \cite{hutchinson1994nonparametric}  use a radial basis function neural network for delta-hedging.
Their network is trained to Black-Scholes prices, using the time-to-maturity and  the moneyness as input variables, without 'shape constraints', i.e. constraints on the derivatives of the learned pricing function. They use the hedging performance of the ensuing delta-hedging strategy as a performance criterion.
\cite{garcia2000pricing} and \cite{genccay2001pricing} improve the stability of the previous approach by adding the outputs of two such neural networks, weighted by respective moneyness and  time-to-maturity functionals.
 \cite{dugas2009incorporating}
 introduce the first neural network
architecture guaranteeing arbitrage-free vanilla option pricing on out-of-sample contracts. In their setup, no-arbitrage is achieved through the choice of special architectures, an approach we subsequently refer to as 'hard constraints'.

However, \cite{ackerer2019deep} show that the corresponding hard constrained networks are very difficult  to train
in practice, leading to unacceptably large errors in price estimation. Instead, they advocate the learning of the implied volatility (rather than the prices) by a standard feedforward neural network with 'soft-constraints', i.e. regularization, which penalizes calendar spread and butterfly arbitrages\footnote{The call and put prices must also be decreasing and increasing by strike respectively.}. Regularization tends to reduce static arbitrage violation on the training set but does not exclude violation on the testing set. This is a by product of using stochastic gradient descent. Unlike interior point methods, which use barrier functions to avoid leaving the feasible set (but are not applicable to neural networks), stochastic gradient descent does not ensure saturation of the penalization (see \citep{marquez2017imposing}).


We propose simple neural network architectures and training methodology which satisfy these shape constraints. Moreover, in contrast to \citep{dugas2009incorporating} and following \citep{ackerer2019deep},  we also explore soft constraints alternatives to hard constraints in the network configuration, due to the loss of representation power of the latter. However,
our networks are trained to prices, versus implied volatilities in \citep{ackerer2019deep}. The closest reference to our work is \cite{itkin2019deep}, who introduces penalty functions to enforce the positivity of the first and second derivatives w.r.t. maturity and strike respectively in addition to the negativity of the first derivative w.r.t strike. Our main contribution with respect to the latter
is 
%
the extraction of a non-parametric representation of the local volatility surface, intrinsically useful for exotic option pricing, but which also serves as a penalization device in our soft constraints approach. The title of the paper emphasizes this difference. In fact, in a single optimization, our approach jointly yields a shape-constrained valuation and the local volatility surface. 
 Using price interpolation, we shall use the Dupire formula to derive a local volatility surface. As we explain later in the paper, such a local volatility surface shall in fact be jointly derived and, at the same time, further regularized.

An additional contribution with respect to
\cite{itkin2019deep,ackerer2019deep} is that these authors only use synthetic data with several thousands of prices, whereas we use real data.
The latter is significant as it is much more representative of the methodology in practice, where noise and a limited number of observations are important aspects of the price interpolation problem.

\begin{figure}[H]
\centering
  \includegraphics[width=0.8\linewidth]{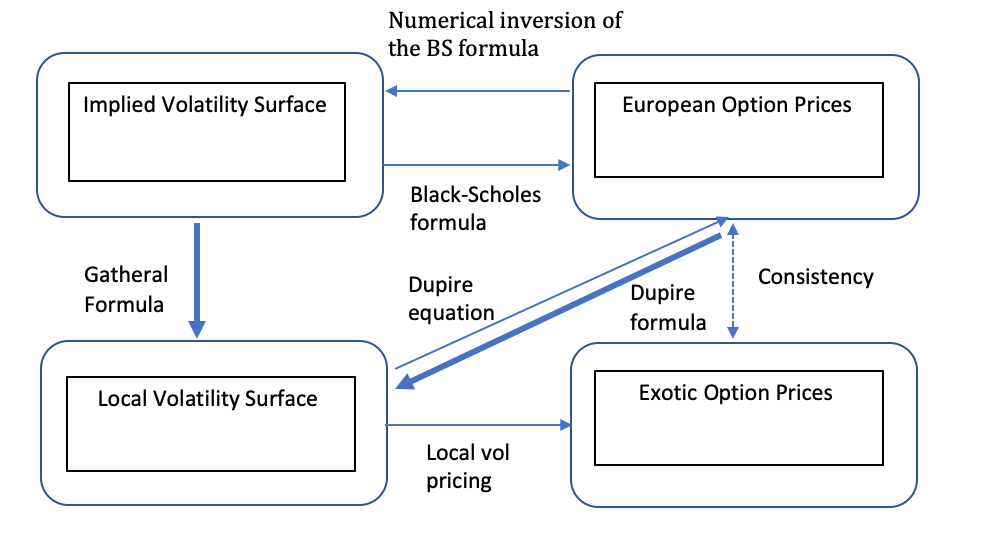} 
\caption{\textit{Mathematical connections between option prices, implied, and local volatility, and the goal of this paper, namely to use the Dupire formula with deep neural networks to jointly approximate the vanilla price and local volatility surfaces.}}
\label{fig:overview}
\end{figure}

An alternative machine learning approach to local volatility calibration is to use the 
\cite{gatheral2011volatility} formula (1.10) to extract the local volatility surface from the Black-Scholes implied volatilities corresponding to the market prices. Figure \ref{fig:overview} recasts the two approaches in the general option pricing perspective. The second approach
will be considered in a forthcoming paper.

\section{Problem Statement}\label{sect:surrogate}

We consider European vanilla option prices on a stock or index $S$.
We assume that a deterministic short interest rate term structure $r(t)$ of the corresponding economy has been bootstrapped from the its zero coupon curve, and that a term structure
of deterministic continuous-dividend-yields $q(t)$ on $S$ has then been extracted from the prices of the forward contracts on $S$. The assumption of deterministic rates and dividends is for consistency with local volatility models, in the perspective, later in the paper, of
numerical experiments on equity options (using  \cite{crepey2002calibration} as a benchmark).

Without restriction given the call-put parity relationship, we only consider put option prices. We denote by $P^{\star}(T,K)$ the market price of the
put option with maturity $T$ and strike $K$ on $S$,
observed for a finite number of pairs $(T,K)$ at a given day.

Our goal is to construct a nonarbitrable put price surface $P:\mathbb{R}_+\times \mathbb{R}_+\rightarrow\mathbb{R}_+$ in $\mathcal{C}^{1,2}((0,+\infty)\times \mathbb{R}_+^\star)\cap\mathcal{C}^{0}(\mathbb{R}_+\times \mathbb{R}_+^\star)$, interpolating $P^\star$ up to some error term.
As is well known,
the corresponding local volatility surface, say $\sigma(\cdot ,\cdot)$, is given by 
the 
\citep{Dupire1994} formula $$
 \frac{\sigma^2(T,K)}{2}
=  
\frac{ \partial_T P(T,K) + (r-q)K\partial_K P(T,K) + qP(T,K)}{K^2 \partial_{K^2}^2 P(T,K)}.
$$ 
In terms of $\thep(T,k) = \exp{(\int_{0}^{T} q(t) dt)} P(T,K),$
where $k= K \exp{(-\int_{0}^{T} (r(t) - q(t)) dt)},$
the formula reads (see Appendix \ref{s:chvar} for a derivation)
\begin{equation}\label{eq:dupk}
\frac{\sigma^2(T,K)}{2} = dup (T,k):= \frac{ \partial_T \thep(T,k)}{k^2 \partial_{k^2}^2 \thep(T,k)}.
\end{equation}  
 We then learn the modified market prices
$\thep^\star=\thep^\star(T,k)$. 
Given a rectangular domain of interest in maturity and strike,
we rescale further the inputs, $T'=(T - min(T))/(max(T)-min(T))$ and $k'=(k - min(k))/(max(k)-min(k)$, so that the domain of the pricing map is $\Omega:=[0,1]^2$. This rescaling avoids any one independent variable dominating over another during the fitting.  For ease of notation, we shall hereon drop the $'$ superscript. 

For the Dupire formula \eqref{eq:dupk} to be meaningful, its output must be nonnegative. This holds, in particular, whenever the interpolating map $\thep$
exhibits nonnegative derivatives w.r.t.~T and second derivative w.r.t.~k, i.e.
\begin{equation}\label{e:na}
\partial_T \thep(T,k)\ge 0\sp  \partial_{k^2}^2 \thep(T,k)\ge 0 .
\end{equation}
In both networks considered in the paper, these derivatives are available analytically via the neural network automatic differentiation capability.
Hard or soft 
constraints can be used to enforce these shape properties, exactly in the case of hard constraints and approximately (via regularization) in the case of soft constraints. More generally,
 see  \cite[Theorem 2.1]{roper2010arbitrage} for a full and detailed statement of the static non-arbitrage relationships conditions 
 on European vanilla call (easily transposable to put) option prices, also including, in particular, an initial condition at $T=0$ given by the option payoffs. This initial payoff condition will also be incorporated to our learning schemes, in a way described in Section \ref{sect:exper}.

\section{Shape Preserving Neural Networks}\label{sect:shape}
We consider 
parameterized maps $p=p_{\mathbf{W},\mathbf{b}}$
$$(T,k)\ni\mathbb{R}_{+}^2 \stackrel{\thep}{\longrightarrow}  \thep_{\mathbf{W},\mathbf{b}}(T,k)  \in \mathbb{R}_+,$$  
given as deep neural networks with two hidden layers. As detailed in \cite{GoodfellowBengioCourville2017}, these
take the form of a composition of simpler functions:
\begin{eqnarray*}
%
&&\thep_{\mathbf{W},\mathbf{b}}(x) =
f^{(3)}_{W^{(3)},b^{(3)}}\circ f^{(2)}_{W^{(2)},b^{(2)}} \circ f^{(1)}_{W^{(1)},b^{(1)}}(x),
\end{eqnarray*}
where $$\mathbf{W} =(W^{(1)},W^{(2)},W^{(3)})\mbox{ and } \mathbf{b}=(b^{(1)},b^{(2)},b^{(3)})$$ 
are weight matrices and bias vectors, and the $f^{(l)} :=\varsigma^{(l)} ( W^{(l)} x + b^{(l)} ) $ are semi-affine, for nondecreasing activation functions
$\varsigma^{(l)} $ 
applied to their (vector-valued) argument componentwise.
Any weight matrix $W^{(\ell)}\in \mathbb{R}^{m\times n}$ can be expressed as an $n$ column $W^{(\ell)}=[\mathbf{w}^{(\ell)}_{1},\dots, \mathbf{w}^{(\ell)}_{n}]$ of $m$-vectors, {for successively chained pairs $(n,m)$ of
dimensions varying with $l=1, 2, 3$, starting from $n=2$, the number of inputs, for $l=1$, and ending up with $m=1$, the number of outputs, for $l=3$.} 

\subsection{Hard Constraints Approach}\label{e:hard}
{In the hard constraints case,}
our
network is sparsely connected in the sense that, with $x=(T,k)$ as above,
\begin{eqnarray*}
f^{(1)}_{W^{(1)},b^{(1)}}(x) = [f^{(1,T)}_{W^{(1,T)}  ,b^{(1,T)}}(T),f^{(1,k)}_{W^{(1,k)},b^{(1,k)}} (k)],
\end{eqnarray*}  
where
$W^{(1,T)},b^{(1,T)}$ and $W^{(1,k)},b^{(1,k)}$ correspond to parameters of sub-graphs for each input, and 
$$f^{(1,T)}(T):=\varsigma^{(1,T)} (W^{(1,T)}T + b^{(1,T)})\sp f^{(1,k)}(k):=\varsigma^{(1,k)} (W^{(1,k)}k + b^{(1,k)}).$$

To impose the shape constraints relevant for put options, it is then enough to
restrict ourselves to  nonnegative weights, and to convex (and nondecreasing) activation functions, namely
$$\mbox{softplus}(x):= \ln(1+e^{x}),$$
except for $\varsigma^{(1,T)}$, which will be taken as an S-shaped sigmoid {$(1+e^{-x})^{-1}$}. Imposing non-negative constraints on weights can be achieved in back-propagation using projection functions applied to each weight after each gradient update.

Hence, the network is convex and nondecreasing in $k$, as a composition (restricted to the $k$ variable) of
convex and nondecreasing functions  of $k$.
In $T$, the network is nondecreasing, but not necessarily convex, because the activation function for the maturity subnetwork hidden layer is not required to be convex - in fact, we choose a sigmoid function.  

\begin{figure}[htbp!]
\centering
\input{figures/nn2_many}

\caption{\textit{A shape preserving (sparse) feed forward architecture with one hidden layer containing 10 hidden nodes. The first input variable, $T$, is only connected to the  5 left most hidden nodes and the second input variable, $k$, is only connected to the 5 right most hidden nodes.}}
\label{fig:architecture} 
\end{figure}
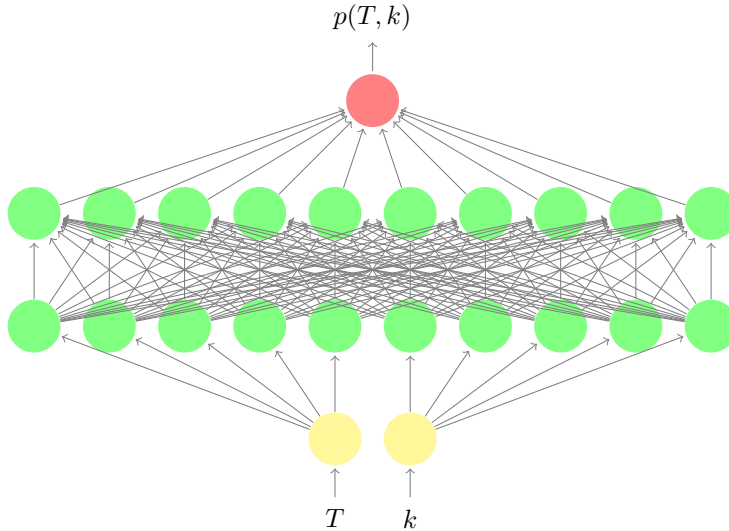

Figure \ref{fig:architecture} illustrates the shape preserving feed forward architecture with two hidden layers containing 10 hidden nodes. For avoidance of doubt, the figure is not representative of the number of hidden neurons 
used in our experiments. However, the connectivity is representative. The first input variable, $T$, is only connected to the first 5 hidden nodes and the second input variable, $k$, is only connected to the last 5 hidden nodes. Effectively, two sub-networks have been created where no information from the input layer crosses the sub-networks until the second hidden layer. In other words, each sub-network is a function of only one input variable. This property is the key to imposing different hard shape constraints w.r.t. each input variable.

\subsection{Soft Constraints Approach}

However,
Theorem 4.1 in \cite{Ohn_2019}, which is stated for the reader's convenience in Appendix \ref{sect:approx}, provides support for our observation, presented in Section \ref{sect:numerical}, that sparsening the network (i.e. splitting) increases the approximation error. 
Hence, in what follows, we also consider the so called soft constraints approach using a fully connected network, 
where the static no arbitrage 
conditions 
\eqref{e:na} are favored by penalization, as opposed to imposed to hold exactly in the previous hard constraint approach.

Note that only the ``hard constraints'' approach theoretically guarantees that the associated Dupire formula
\eqref{eq:dupk}
 returns a positive function. While soft constraints reduce the risk of static arbitrage in the sense of mismatch between model and  market prices, they do not however fully prevent arbitrages in the sense of violations of the shape conditions \eqref{e:na} in the predicted price surface, especially far from the grid nodes of the training set.

In particular, the penalties only control the corresponding derivatives at the training points. 
Compliance with the no-arbitrage constraints on the majority of the points in the test set is due only to the regularity of these derivatives.
This is not a novel idea. \cite{aubinfrankowski2020hard} exploit RKHS regularity to ensure  conditions on derivatives in a hard constraint manner with a second order cone optimization.
They add a margin to the penalizations so that these derivative conditions are ensured over a
targeted
neighbourhood of training points.
In our case we do not add such a margin to our penalizations.
Instead, we add a further half-variance bounds penalization, which both favors even more the shape constraints (without guaranteeing them in theory) and stabilizes the local volatility function calibration, as detailed below.

\section{Numerical Methodology}
 \subsection{Training}
\label{sect:training}
In general, to fit our fully connected or sparse networks to the available option market prices at a given 
time,
we solve a loss minimization problem of the following form (with $\lambda=0$ in the non-penalized cases), using observations $\{x_i=(T_i,k_i),\thep^\star(x_i)\}_{i=1}^n$ of $n$ maturity-strike pairs and the corresponding market
put prices: 
\be \label{e:loss} 
\min_{\mathbf{W},\mathbf{b}}\;\; 
\frac{1}{n} 
\sum_{i=1}^n 
\Big( 
| \thep^{\star}(x_i ) -\thep(x_i )| +  \lambda^{\sf T}\phi(x_i ) \Big).
\ee
Here $p=\thep_{\mathbf{W},\mathbf{b}}$, $\phi=\phi_{\mathbf{W},\mathbf{b}}$ is a regularization penalty vector 
\begin{eqnarray*}
\phi :=[
(\partial_T \thep)^-,   ( \partial^2_{k^2} \thep )^-,
 (dup -\overline{a})^+
  +  (dup -\underline{a} )^-
],
\end{eqnarray*} 
and $dup$ is related to $\thep$ through \eqref{eq:dupk}. The choice to measure the error $\thep^{\star} -\thep$ under the $L_1$ norm, rather than $L_2$ norm, in \eqref{e:loss} is motivated by a need to avoid allocating too much weight to the deepest in-the-money options. Note that \cite{ackerer2019deep} consider a combination of $L_1$ and $L_2$ norms. In a separate experiment, not reported here, we additionally investigated using the market convention of vega weighted option prices, albeit to no effect beyond simply using $L_1$ regularization.

The loss function is non-convex, possessing many local minima and it is generally difficult to find a global minimum. The first two elements in the penalty vector favor the shape conditions \eqref{e:na} and the third element favors lower and upper bounds 
$\underline{a}$ and 
$\overline{a}$
  on the estimated half-variance, $dup$, where constants (which could also be functions of time) $0<\underline{a}<\overline{a}$ respectively denote desired lower and upper bounds on the surface (at each point in time).
Of course, as soon as penalizations are effectively used (i.e. for $\lambda\neq 0$), 
a further difficulty, typically involving grid search, is the need to determine suitable values of the corresponding ``Lagrange multipliers" 
\be \label{e:lambdas} \lambda=(\lambda_1,\lambda_2,\lambda_3)\in \mathbb{R}_+^3,\ee ensuring the right balance between fit to the market prices and the targeted constraints.

\subsection{Experimental Design}
 \label{sect:exper}

As a benchmark, reference method for assessing the performance of our neural network approaches, we use the
Tikhonov regularization approach surveyed Section 9.1 of \cite{cre}, i.e.
nonlinear least square minimization of the squared distance to market prices plus a penalisation proportional to the $H^1$ squared norm of the local volatility function over the (time, space) surface (or, equivalently, to the $L^2$ norm of the gradient of the local volatility).
Our motivation for this choice as a benchmark is, first, the theoretical, mathematical justification for this method provided by Theorems 6.2 and 6.3 in \citep{crepey2003calibration}. Second,
it is price (as opposed to implied volatility) based, which makes it at par with our focus on {\em price based} neural network  local volatility calibration schemes in this paper. Third, it is non parametric ('model free' in this sense), like our 
neural network schemes again, and as opposed to various parameterizations such as SABR or SSVI that are used as standard in various segments of the industry, but come without theoretical justification for robustness, are restricted to specific industry segments on which they play the role of a market consensus, and are all implied volatility based. Fourth, 
an efficient numerical implementation of the Tikhonov method (as we call it for brevity hereafter), already put to the test of hundreds of real datasets in the context of 
\citep{crepey2004delta}, is available through \citep{crepey2002calibration}. 
Fifth, this method is itself benchmarked to other (spline interpolation and constrained stochastic control) approaches Section 7 of \citep{crepey2002calibration}. 
 
\begin{figure}[htbp!]
\centering
\begin{tabular}{cc}
 \includegraphics[width=1.0\linewidth]{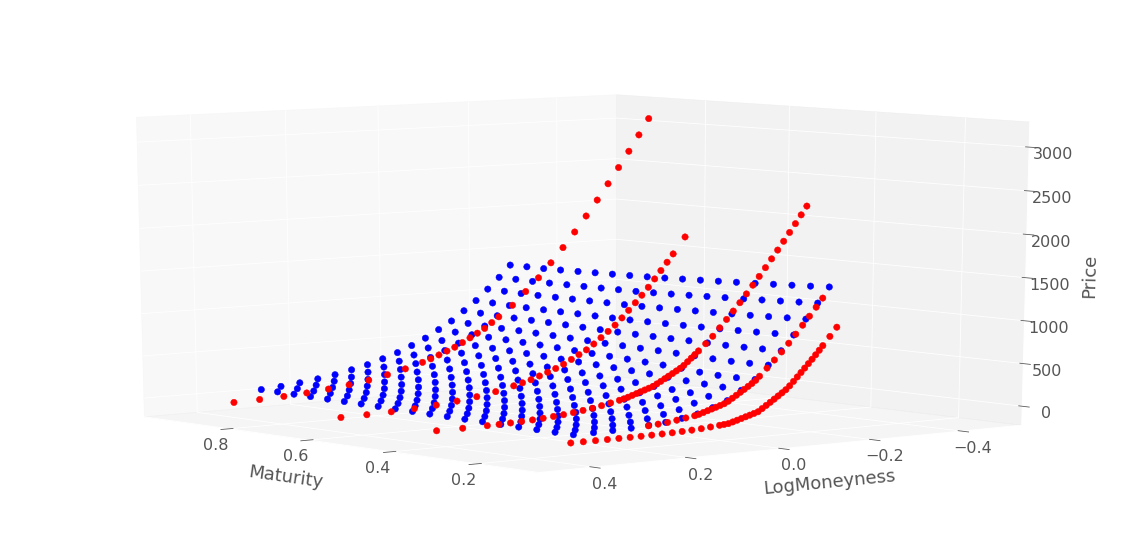} 
   \end{tabular}
   \caption{\textit{DAX put prices from
   training grid (red points) and testing grid (blue points), 8 Aug 2001.}}
\label{fig:SurfacePricesOriginal}
\end{figure}

\begin{figure}[htbp!]
\centering 
\begin{tabular}{cc} 
 \includegraphics[width=1.0\linewidth]{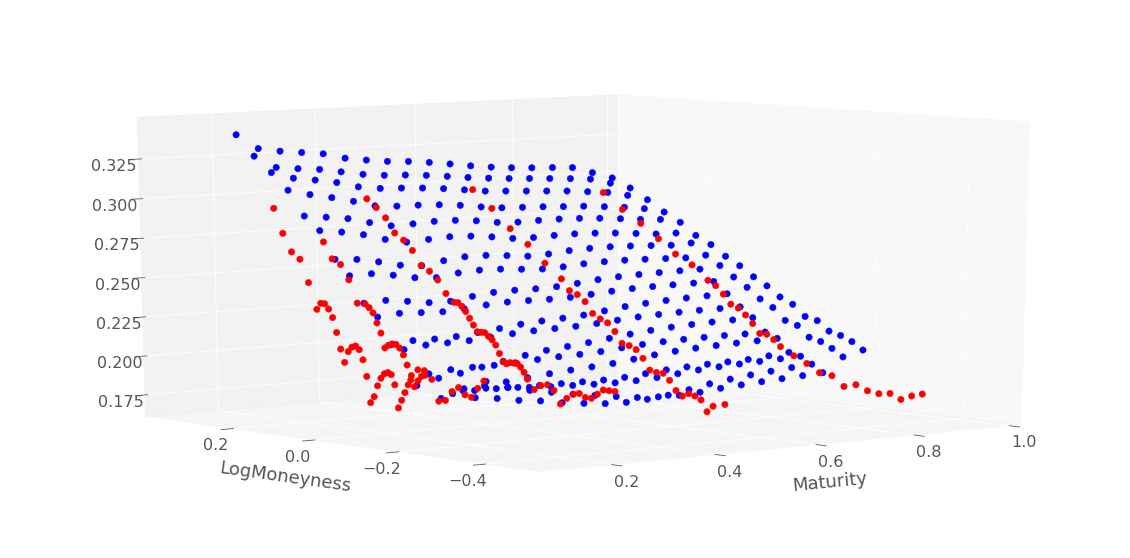} 
   \end{tabular}
   \caption{\textit{Same as Figure  \ref{fig:SurfacePricesOriginal} in  implied volatility scale.}}
\label{fig:SurfaceVolsOriginal}
\end{figure}

Our training sets are prepared using daily datasets of DAX index European vanilla options of different available strikes and maturities, listed on the $7^{th}$, $8^{th}$ (by default below), and $9^{th}$, August 2001, i.e. same datasets as already used in previous work  \citep {crepey2002calibration,crepey2004delta}, for benchmarking purposes. The corresponding values of the underlying are $S=5752.51, 5614.51
$ and $5512.28$. The associated interest rate and dividend yield curves are constructed from zero-coupon and forward curves, themselves obtained from quotations of standard fixed income linear instruments and from call/put parity applied to the option market prices.
 Each training set is composed of about
 200 option market prices plus the put payoffs for all strikes present in the training grid. 
 For each day of data
(see e.g. Figures \ref{fig:SurfacePricesOriginal}-\ref{fig:SurfaceVolsOriginal}), 
a test set of about 350 points is generated by computing, thanks to a trinomial tree, option prices for a regular grid of strikes and maturities, in the local volatility model calibrated to the corresponding
training set by the benchmark Tikhonov calibration method of \cite {crepey2002calibration}.

Each network has two hidden layers, each with 200 neurons per hidden layer. Note that \cite {dugas2009incorporating} only uses one hidden layer. Two was found important in practice in our case. All networks are fitted with an ADAM optimizer. In order to achieve the convergence of the training procedure toward a local minimum of the loss criterion, the learning rate is divided by 10 whenever no improvement in the error on the training set is observed during 100 consecutive epochs. The total number of epochs is limited to $ 10,000 $ because of the limited number of market prices. 
Thus we opt for a batch learning with numerous epochs.

After training, a local volatility surface is extracted from the interpolated prices by application of the Dupire formula \eqref{eq:dupk}, leveraging on the availability of the corresponding exact sensitivities, i.e., using automatic algorithmic differentiation (AAD) and not finite differences. This local volatility surface is then compared with the one obtained in \cite {crepey2002calibration}.

Moreover,
we will assess numerically four different combinations of network architectures and optimization criteria, i.e.
\begin{itemize}
\item sparse (i.e. split) network and hard constraints, so $\lambda_1= \lambda_2=0$ in \eqref{e:loss}-\eqref{e:lambdas},
\item sparse network but soft constraints, i.e. ignoring the non-negative weight 
restriction in Section \ref{e:hard}, but using $\lambda_1, \lambda_2 >0$ in \eqref{e:loss}-\eqref{e:lambdas},
\item  dense network and soft constraints, i.e. for $\lambda_1, \lambda_2 >0$ in 
\eqref{e:loss}-\eqref{e:lambdas},
\item  dense network and no shape constraints, i.e. $\lambda_1= \lambda_2=0$ in \eqref{e:loss}-\eqref{e:lambdas}.
\end{itemize}
Moreover, these four configurations will be tried both without  (Section \ref{sect:numerical}) and with (Section \ref{sect:numerical2})  half-variance bounds penalization, i.e. for $\lambda_3 =0$ vs. $\lambda_3 >0$ in 
\eqref{e:loss}-\eqref{e:lambdas}, cases referred to hereafter as without / with Dupire penalization.

In each case the error between the prices of the calibrated model and the market data are evaluated on both the training and an out-of-sample test set.
Unless reported otherwise, all numerical results shown below correspond to test sets. 

All our numerical experiments were run under google colab with 13 Gos of Ram and a dual core CPU of 2.2GHz. 

\section{Numerical Results Without Dupire Penalization}\label{sect:numerical}

Table \ref{tab:myTable1} shows the pricing RMSEs for four different combinations of architecture and optimization criteria
without half-variance bounds, i.e. for $\lambda_3 =0$  
\eqref{e:loss}-\eqref{e:lambdas}.
For the sparse network with hard constraints, we thus have $\lambda=0$. For the sparse and dense networks with soft constraints (i.e. penalization but without the conditions on the weights of Section \ref{sect:shape}), we set $\lambda=[1.0\times 10^5, 1.0\times 10^3,0]$.  

The sparse network with hard constraints
is observed to exhibit significant pricing error, which suggests that this approach is too limited in practice to approach market prices. This conclusion is consistent with \cite{ackerer2019deep}, who choose a soft-constraints approach in the implied volatility approximation (in contrast to our approach which approximates prices).

 \begin{table}[!ht]
 \resizebox{\columnwidth}{!}{%
\begin{tabular}{|l|c|c|c|c|}
\hline
                 & \multicolumn{2}{c|}{Sparse network} & \multicolumn{2}{c|}{Dense network}       \\ \hline
                 & Hard constraints  & Soft constraints  & Soft constraints & No constraints \\ \hline
Training dataset & 28.13             & 6.87              & 2.28             & 2.56                  \\ \hline
Testing dataset  & 28.91             & 4.09              & 3.53             & 3.77                  \\ \hline
Indicative training  times & 200s           & 400s        & 200s         & 120s      \\ \hline
\end{tabular}
}
\caption{\textit{Pricing RMSE (absolute pricing errors) and training times without Dupire penalization.}}
\label{tab:myTable1}
\end{table}

\begin{figure}[htbp!]
\centering
\begin{tabular}{cc}
  \includegraphics[width=\linewidth]{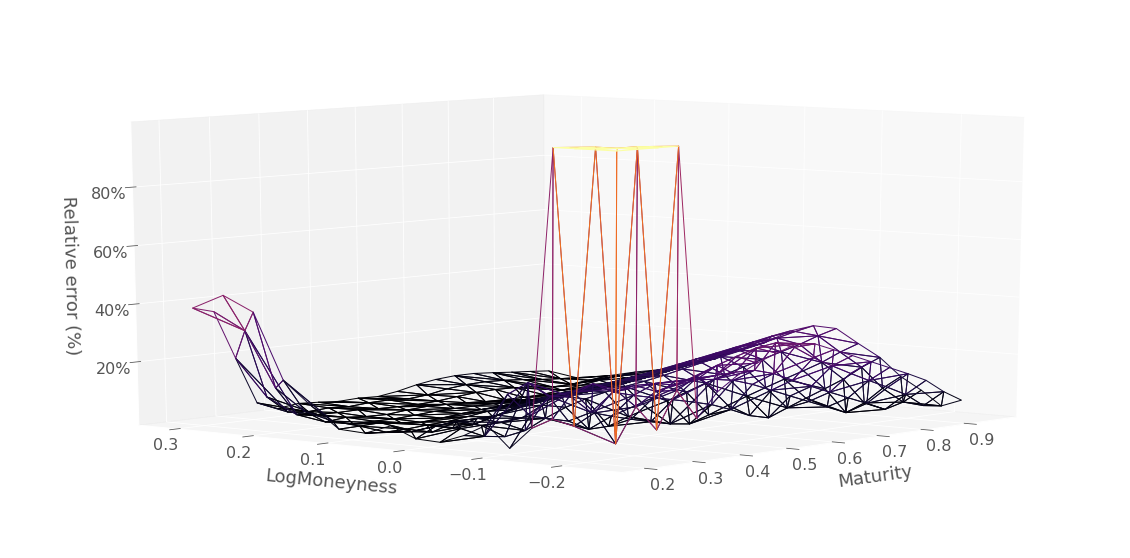} 
   \\
 \includegraphics[width=\linewidth]{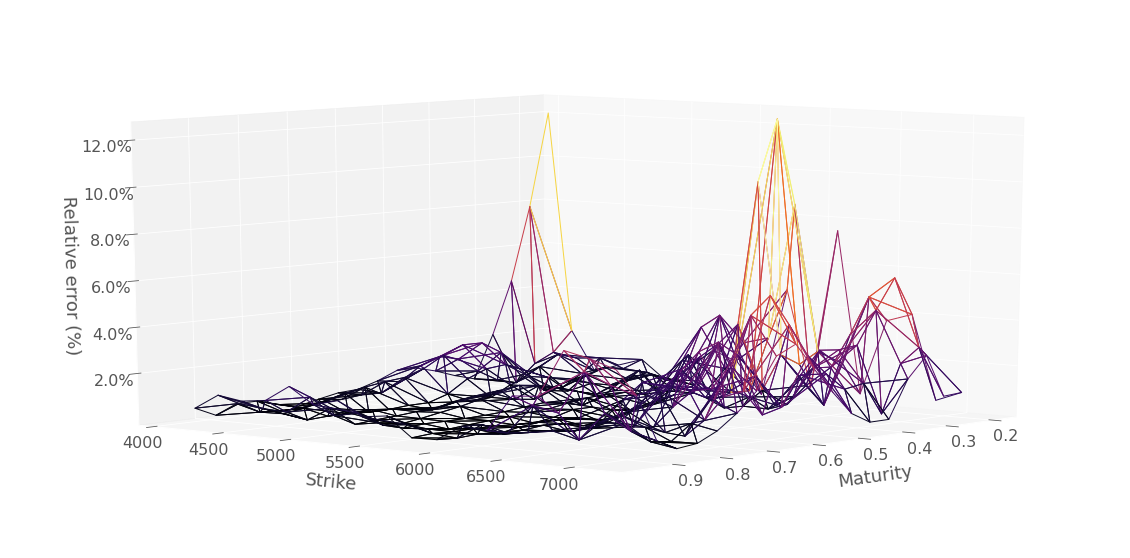} 
   \end{tabular}
   \caption{\textit{Percentage relative error in the implied volatilities using (top) hard constraints (bottom) dense networks with soft constraints.}}
\label{fig:errorHardImpli}
\end{figure}

Figure \ref{fig:errorHardImpli}
compares the percentage errors in implied volatilities 
using the sparse network with hard constraints and the dense network with soft constraints approaches, corresponding to the columns 1 and 3 of Table \ref{tab:myTable1}. Relative errors with hard constraints exceed 10\% on most the training grid oppositely to dense network with soft constraints. This confirms that the error levels of the hard constraints approach are too high to imagine a practical use of this approach: the corresponding model would be immediately arbitrable in relation to the market. Those of the soft constraint approach are much more acceptable, with high errors confined to short maturities or far from the money, i.e. in the region where prices provide little information on volatility.

Table 
\ref{tab:SAUnregularized}
shows the fraction of points in the neural network price surface which violate the static arbitrage conditions.
The table compares the same four methods listed in Table \ref{tab:myTable2} applied to training and testing sets. We recall that, in theory, only the sparse network with hard constraints guarantees zero arbitrages. However, we observe that the inclusion of soft constraints reduces the number of arbitrage constraints on the training set when compared with no constraints. The trend is less pronounced for the test set. But in the absence of hard constraints, the effect of adding soft constraints is always preferable than excluding them entirely.

\begin{table}[!htbp]
 \resizebox{\columnwidth}{!}{%
\begin{tabular}{|l|c|c|c|c|}
\hline
                 & \multicolumn{2}{c|}{Sparse network} & \multicolumn{2}{c|}{Dense network}       \\ \hline
                 & Hard constraints  & Soft constraints  & Soft constraints & No constraints \\ \hline
Training dataset &        0      &      1/254        &        0     &         63/254         \\ \hline
Testing dataset  &    0          &       2/360      &        0      &      44/360             \\ \hline
\end{tabular}
}
\caption{\textit{The fraction of static arbitrage violations without Dupire penalization.}}
\label{tab:SAUnregularized}
\end{table}

 \section{Numerical Results With Dupire Penalization}
 \label{sect:numerical2}
 
\begin{figure}[htbp!]
\centering 
\begin{tabular}{cc} 
 \includegraphics[width=1.0\linewidth]{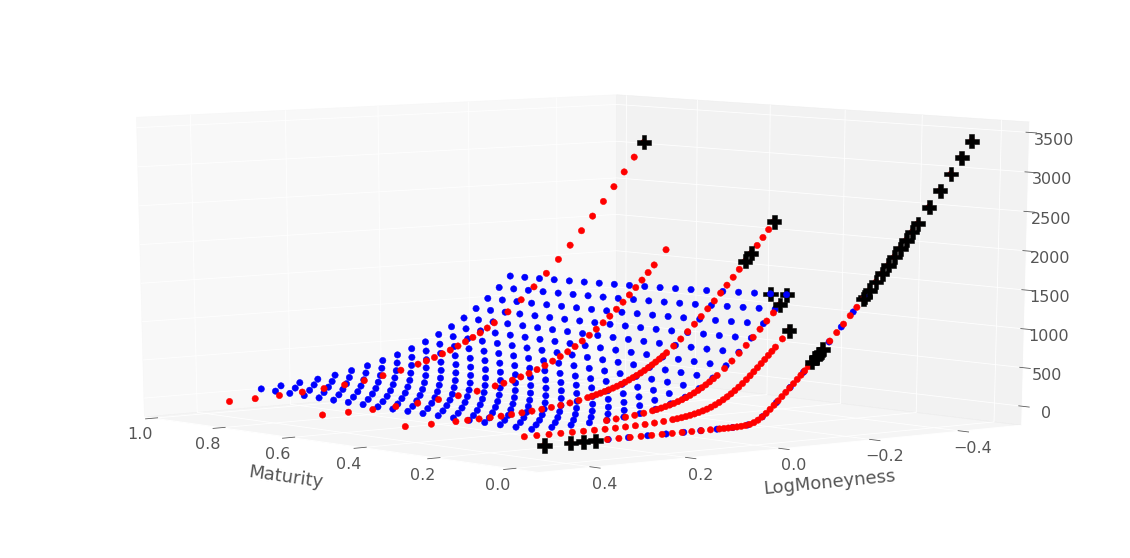} 
   \end{tabular}
   \caption{\textit{Location of the violations, denoted by black crosses, corresponding to the right column in Table \protect\ref{tab:SA}.}}
\label{fig:viola}
\end{figure}

We now introduce half-variance bounds into the penalization to improve the overall fit in prices and stabilize the local volatility surface. Table \ref{tab:myTable2} shows the RMSEs in absolute pricing resulting from repeating the same set of experiments reported in Table \ref{tab:myTable1}, but with the half-variance bounds included in the penalization.  For the sparse network with hard constraints, we set $\lambda=[0,0,10]$ and choose $\underline{a}=0.05^2/2$ and $\overline{a}=0.4^2/2$. For the sparse and dense networks with soft constraints, we set $\lambda=[1.0\times 10^5, 1.0\times 10^3,10]$. Compared to Table \ref{tab:myTable1}, we observe improvement in the test error for the hard and soft constraints approaches when including the additional local volatility penalty term. 
Table \ref{tab:SA} is the analog of Table \ref{tab:SAUnregularized}, with similar conclusions. Note that, here as above, the  arbitrage
opportunities that arise are not only very few (except in the unconstrained case), but also very far from the money and, in fact, mainly regard the learning of the payoff function, corresponding to the horizon $T=0$. See for instance Figure \ref{fig:viola} for the location of the violations that arise in the unconstrained case with Dupire penalization. Hence such apparent 'arbitrage opportunities' cannot necessarily be monetised once liquidity is accounted for.

\begin{table}[!ht]
 \resizebox{\columnwidth}{!}{%
\begin{tabular}{|l|c|c|c|c|}
\hline
                 & \multicolumn{2}{c|}{Sparse network} & \multicolumn{2}{c|}{Dense network}       \\ \hline
                 & Hard constraints  & Soft constraints  & Soft constraints & No constraints \\ \hline
Training dataset & 28.04             & 3.44              & 2.48             & 3.48                  \\ \hline
Testing dataset  & 27.07             & 3.33              & 3.36             & 4.31                  \\ \hline
Indicative training  times &   400s          & 600s           &   300s            &  250s            \\ \hline
\end{tabular}
}
\caption{\textit{Price RMSE (absolute pricing errors) and training times with Dupire penalization.}}
\label{tab:myTable2}
\end{table}

\begin{table}[!htbp]
 \resizebox{\columnwidth}{!}{%
\begin{tabular}{|l|c|c|c|c|}
\hline
                 & \multicolumn{2}{c|}{Sparse network} & \multicolumn{2}{c|}{Dense network}       \\ \hline
                 & Hard constraints  & Soft constraints  & Soft constraints & No constraints \\ \hline
Training dataset &        0      &      0        &        0     &         30/254         \\ \hline
Testing dataset  &    0          &       2/360       &        0      &      5/360             \\ \hline
\end{tabular}
}
\caption{\textit{The fraction of static arbitrage violations with Dupire penalization.}}
\label{tab:SA}
\end{table}

\begin{figure}[htbp!]
\centering
\begin{tabular}{cc}
  \includegraphics[width=1.0\linewidth]{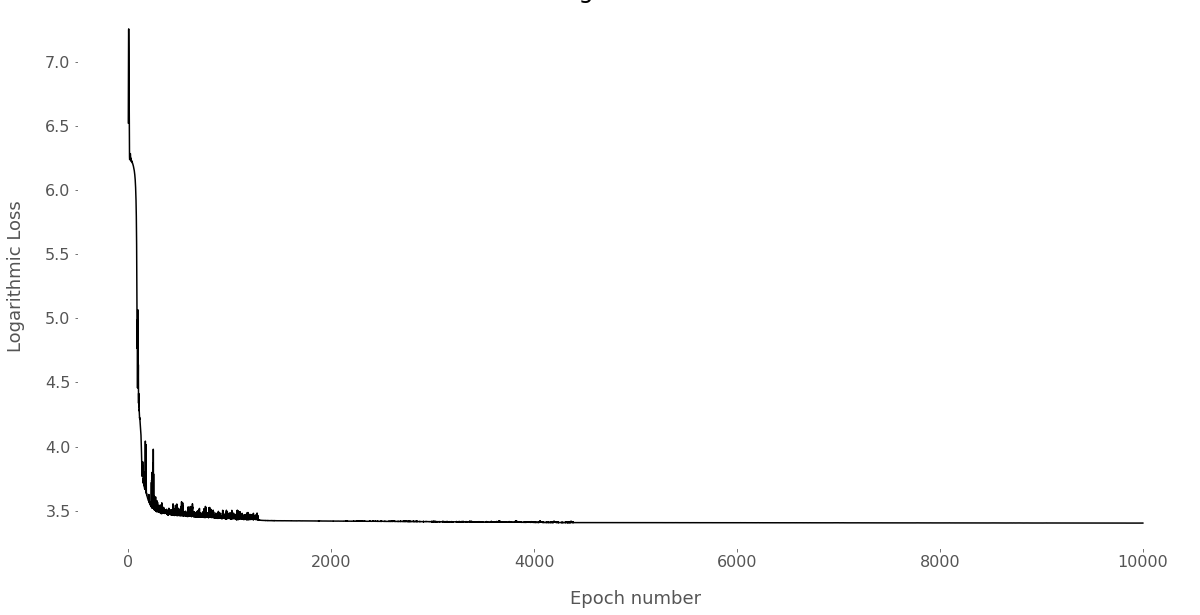} \\
 \includegraphics[width=1.0\linewidth]{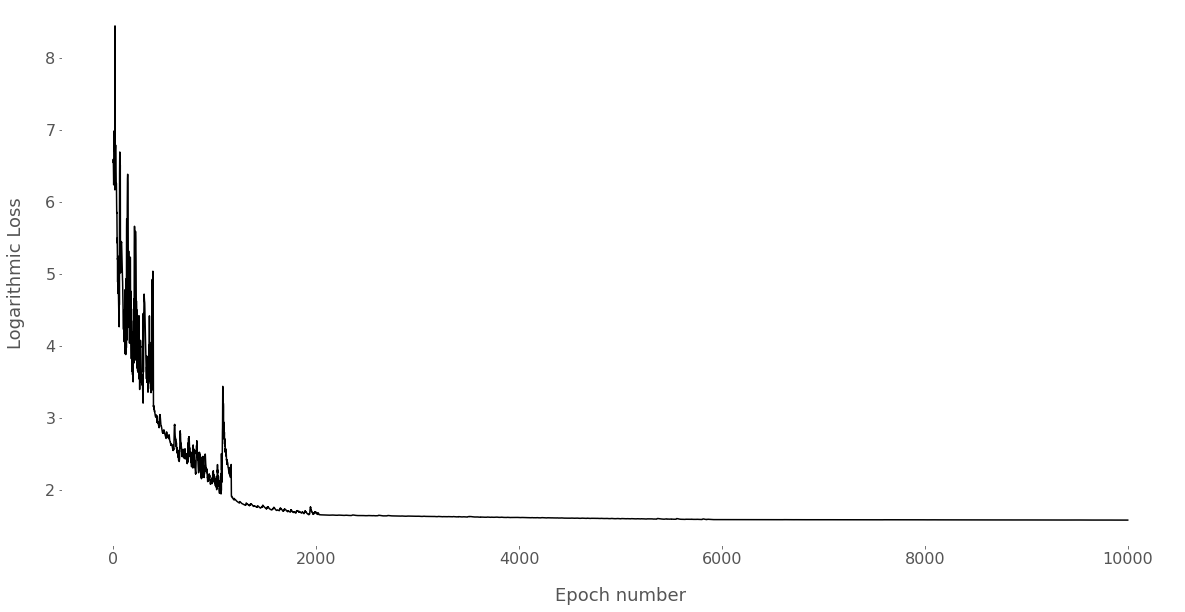} 
   \end{tabular}
   \caption{\textit{Logarithmic RMSE through epochs (top) hard constraints (bottom) dense networks with soft constraints.}}
\label{fig:LossPlots}
\end{figure}
\begin{figure}[htbp!]
\centering 
\begin{tabular}{cc}
 \includegraphics[width=\linewidth]{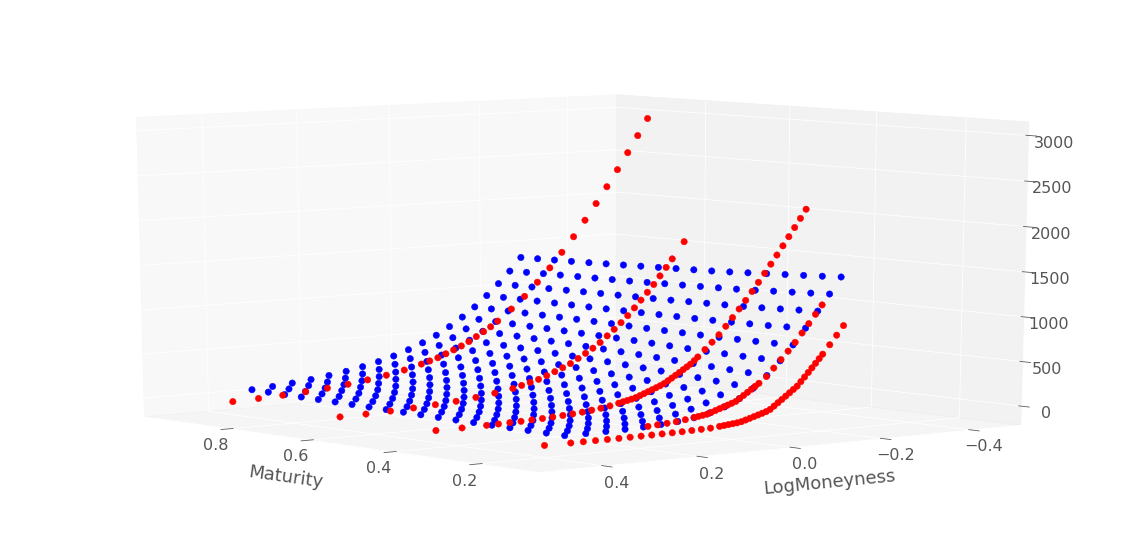} 
   \end{tabular}
   \caption{\textit{Put prices from
   training grid (red points) and NN predicted prices at testing grid nodes (blue points), DAX 8 Aug 2001.}}
\label{fig:SurfacePricesNNPredict}
\end{figure}

\begin{figure}[htbp!]
\centering 
\begin{tabular}{cc}
  \includegraphics[width=\linewidth]{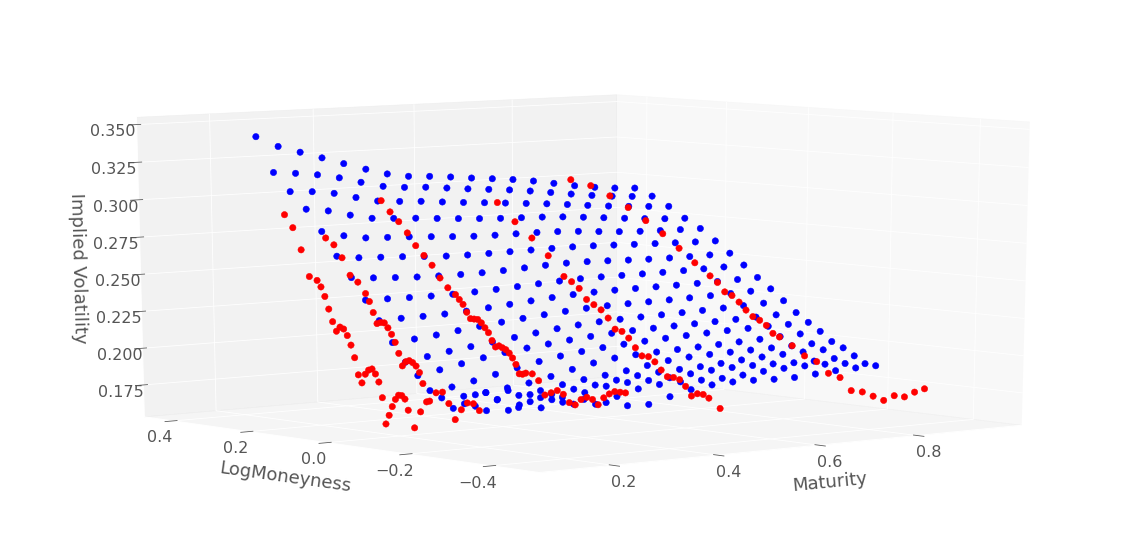} 
   \end{tabular}
   \caption{Same as Figure \ref{fig:SurfacePricesNNPredict}
   in implied volatility scale.}
\label{fig:SurfaceVolsNNPredict}
\end{figure}

%
Figure \ref{fig:SurfacePricesNNPredict} is the analog of
Figure \ref{fig:SurfacePricesOriginal}, with test (i.e. Tikhonov trinomial tree) prices in blue replaced by the prices predicted by the dense network with soft constraints and Dupire penalization.
The (blue) prices predicted by the neural network in Figure \ref{fig:SurfacePricesNNPredict}, and the corresponding implied volatilities in Figure \ref{fig:SurfaceVolsNNPredict}, do not exhibit any visible inter-extrapolation pathologies, they are in fact visually indistinguishable from the respective  (blue) testing prices and implied volatilities of Figures \ref{fig:SurfacePricesOriginal} and \ref{fig:SurfaceVolsOriginal}.  

For completeness, we additionally provide further diagnostic results.  Figure \ref{fig:LossPlots} shows the convergence of the loss function against the number of epochs using either hard constraints or soft constraints. The spikes trigger decays of the learning rates so that 
 the training procedure can converge toward a local minimum of the loss criterion
(cf. Section 
 \ref{sect:exper}).
We observe that the loss function converges to a much smaller value using a dense network with soft constraints and that either approach converge in at most 2000 epochs.

Table \ref{tab:xval} provides some further insight into the effect of architectural parameters, although it is not intended to be an exhaustive study. Here, only the number of units in the hidden layers is varied, while keeping all other parameters except the learning rate fixed, to study the effect on error in the price and implied volatility surfaces. The price RMSE for the testing set primarily provides justification for the choice of 200 hidden units per layer: the RMSE is 3.55. We further observe the effect of reduced pricing error on the implied volatility surface: 0.0036 is the lowest RMSE of the implied volatility test surface across all parameter values.

\begin{table}[!ht]
 \centering
\begin{tabular}{|l|l|c|c| }
\hline
   \# Hidden Units              & Surface & \multicolumn{2}{c|}{RMSE}     \\ \hline
                 & & Training  & Testing \\ \hline
50 &   Price &  3.01        &     3.60                       \\
 & Impl. Vol. & 0.0173 & 0.0046 \\ \hline

100 &   Price &   3.14       &     3.66                        \\
 & Impl. Vol. & 0.0304 & 0.0049 \\ \hline
 
 \bf{200} &   Price &    2.73    & 3.55                           \\
 & Impl. Vol. & 0.0181 & 0.0036 \\ \hline
 
 300 &   Price &    2.84      &    3.88                         \\
 & Impl. Vol. & 0.0180  & 0.0050 \\ \hline
 
 400 &   Price &    2.88      &     3.56                        \\
 & Impl. Vol. & 0.0660  & 0.0798 \\ \hline
\end{tabular}
\caption{\textit{Sensitivity of the errors to the number of hidden units. Note that these results are generated using the dense network with soft constraints and Dupire penalization.}}
\label{tab:xval}
\end{table}

Table \ref{tab:GradAlgo} shows the pricing RMSEs resulting from the application of different stochastic gradient descent algorithms under the soft
constraints approach with dense network and Dupire penalization. ADAM (our choice everywhere else in the paper, cf. the next-to-last column in Table \ref{tab:myTable2}) and RMSProp (root mean square propagation, another well known SGD procedure) exhibit a comparable performance. 
A Nesterov accelerated gradient procedure, with momentum parameter set to 0.9 as standard, obtains much less favorable results. As opposed to
ADAM and RMSProp, Nesterov accelerated momentum does not reduce the learning rate during the optimization.


\begin{table}[!ht]
\centering
\begin{tabular}{|c|c|c|}
\hline
                                                                                       & Train RMSE & Test RMSE \\ \hline
ADAM                                                                                   & 2.48       & 3.36      \\ \hline
Nesterov accelerated gradient & 5.67       & 6.92      \\ \hline
RMSProp                                                                                & 2.76       & 3.66      \\ \hline
\end{tabular}
\caption{Pricing RMSEs corresponding to different stochastic gradient descents (soft
constraints approach with dense network and Dupire penalization). }
\label{tab:GradAlgo}
\end{table}

\section{Robustness}

In this concluding section of the paper, we assess the robustness of the different approaches in terms of, first, the numerical stability of the local volatility function recalibrated across successive calendar days and, second, of a Monte Carlo backtesting repricing error.

\subsection{Numerical Stability Through Recalibration}

\begin{sidewaysfigure}[htbp]
\centering
\begin{subfigure}{.5\textwidth}
  \centering
  \includegraphics[width=\linewidth]{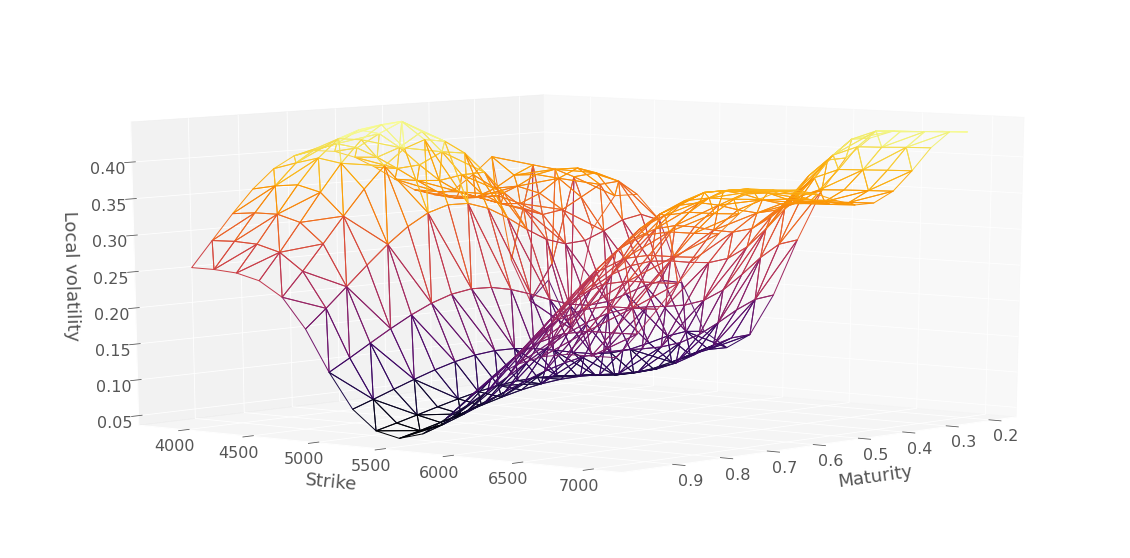}
  \caption{Hard Constraints}
  \label{fig:07082001HardReg}
\end{subfigure}
\begin{subfigure}{.48\textwidth}
  \centering
  \includegraphics[width=1.0\linewidth]{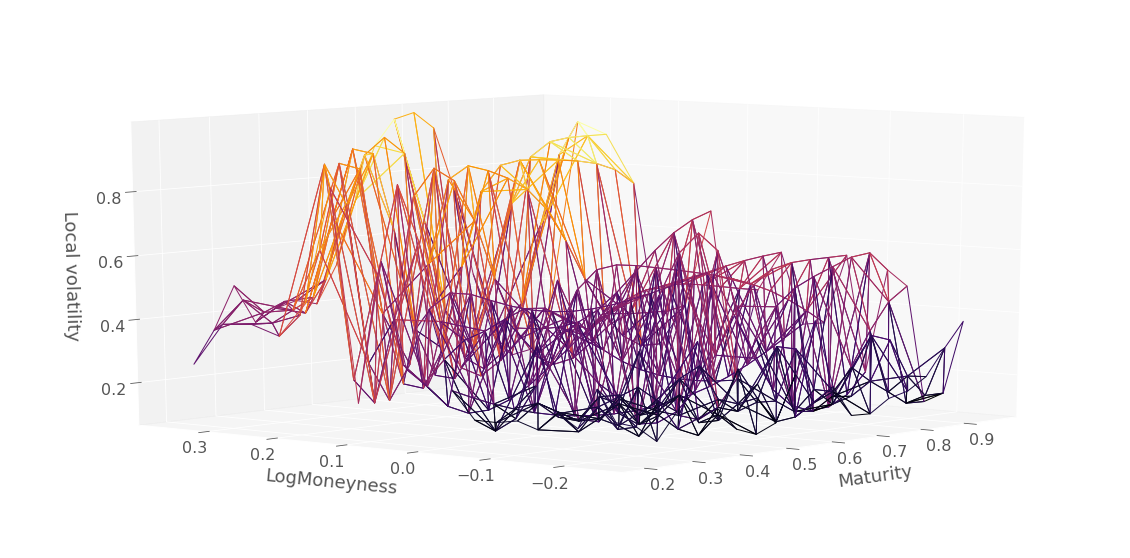}
  \caption{Soft constraints (w/o local vol. constraints)}
  \label{fig:07082001Soft2}
\end{subfigure}

\begin{subfigure}{.5\textwidth}
  \centering
  \includegraphics[width=\linewidth]{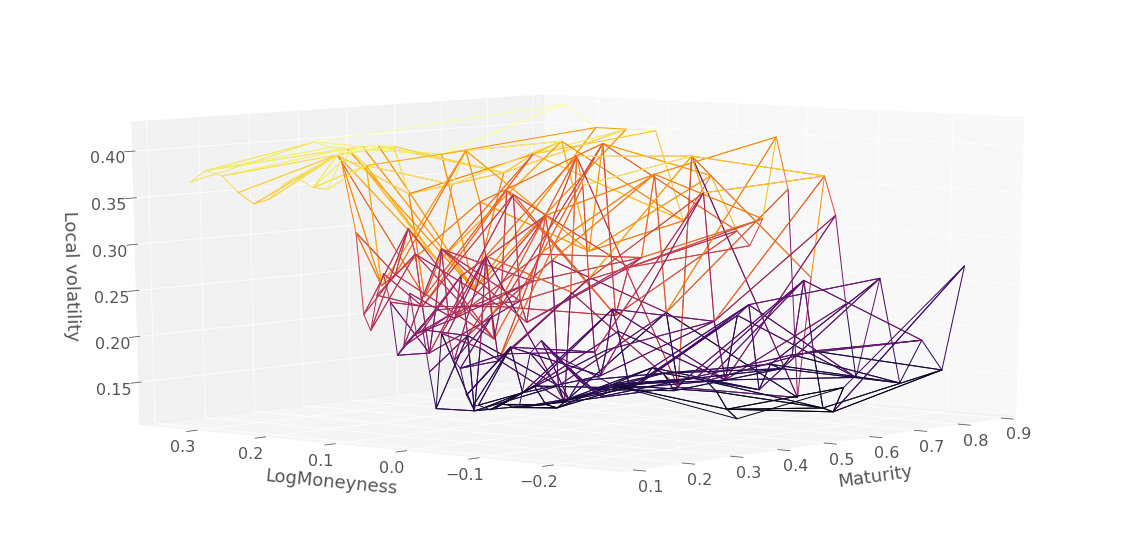}
  \caption{Soft constraints (with local vol. constraints)}
  \label{fig:07082001Soft2bis}
\end{subfigure}
\begin{subfigure}{.48\textwidth}
  \centering
  \includegraphics[width=1.0\linewidth]{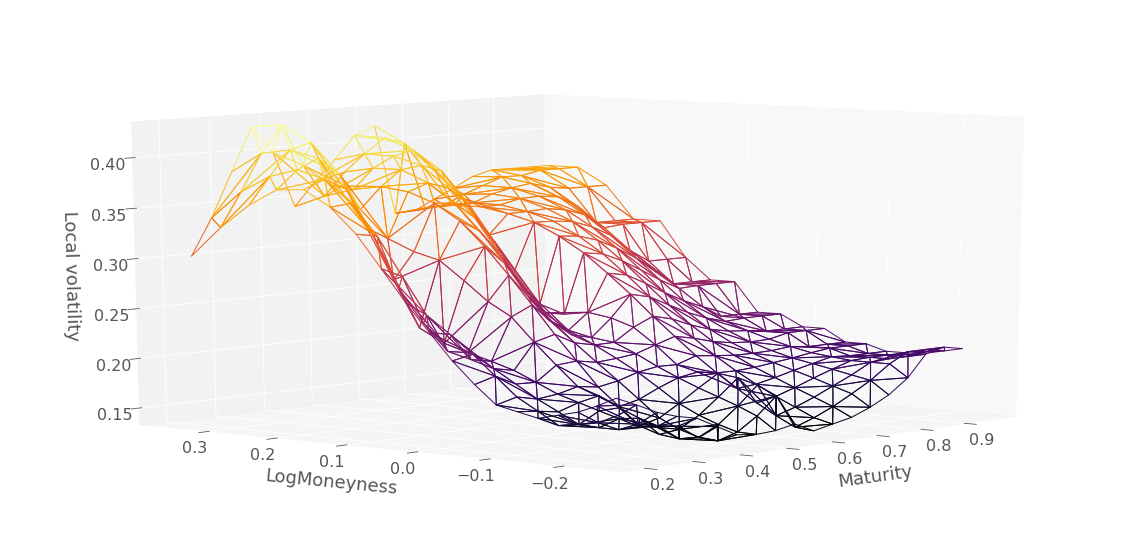}
  \caption{Tikhonov local volatility}
  \label{fig:07082001SoftTrue}
\end{subfigure}%

\caption{\textit{Local volatility for 07/08/2001.}}
\label{fig:testMulti12}
\end{sidewaysfigure}

\begin{sidewaysfigure}[htbp]
\centering
\begin{subfigure}{.48\textwidth}
  \centering
  \includegraphics[width=1.0\linewidth]{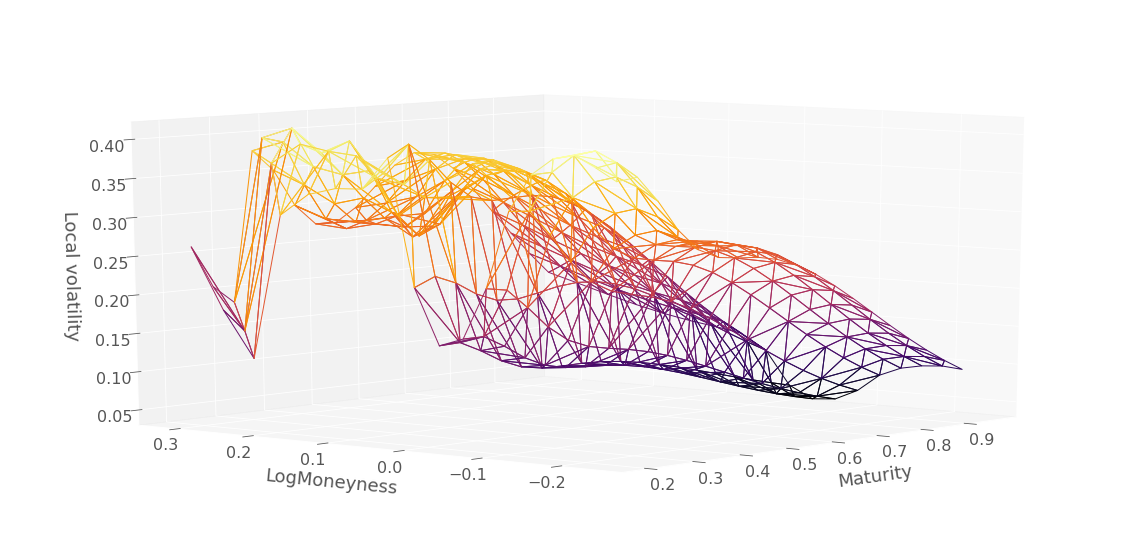}
  \caption{Hard Constraints}
  \label{fig:08082001HardReg}
\end{subfigure}
\begin{subfigure}{.48\textwidth}
  \centering
  \includegraphics[width=1.0\linewidth]{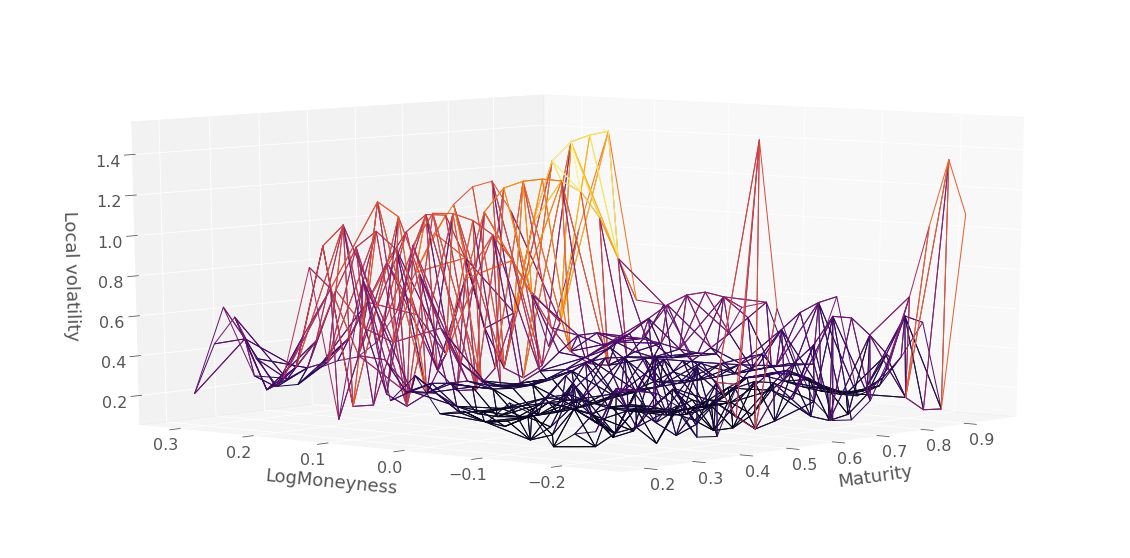}
  \caption{Soft constraints (w/o local vol. constraints)}
  \label{fig:08082001Soft2}
\end{subfigure}%

\begin{subfigure}{.48\textwidth}
  \centering
  \includegraphics[width=1.0\linewidth]{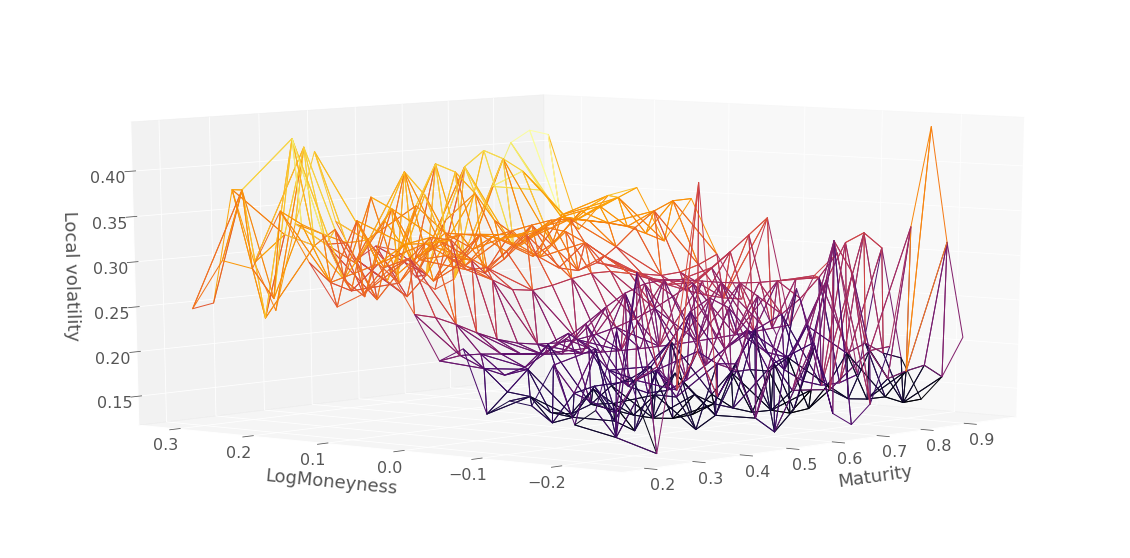}
  \caption{Soft constraints (with local vol. constraints)}
  \label{fig:08082001Soft2bis}
\end{subfigure}%
\begin{subfigure}{.48\textwidth}
  \centering
  \includegraphics[width=1.0\linewidth]{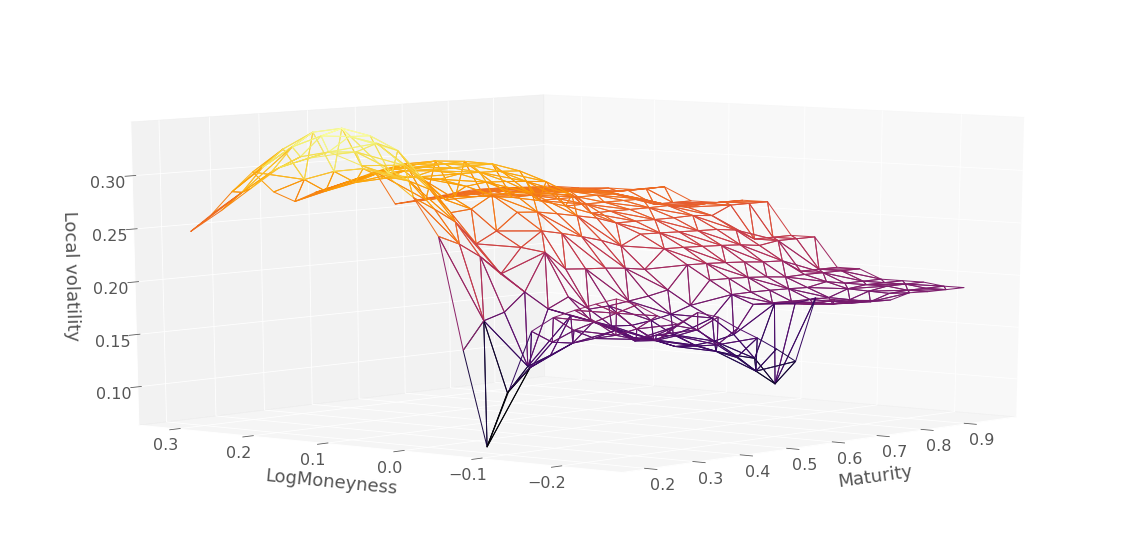}
  \caption{Tikhonov local volatility}
  \label{fig:08082001SoftTrue}
\end{subfigure}%

\caption{\textit{Local volatility for 08/08/2001.}}
\label{fig:testMulti13}
\end{sidewaysfigure}

\begin{sidewaysfigure}[htbp]
\centering

\begin{subfigure}{.48\textwidth}
  \centering
  \includegraphics[width=1.0\linewidth]{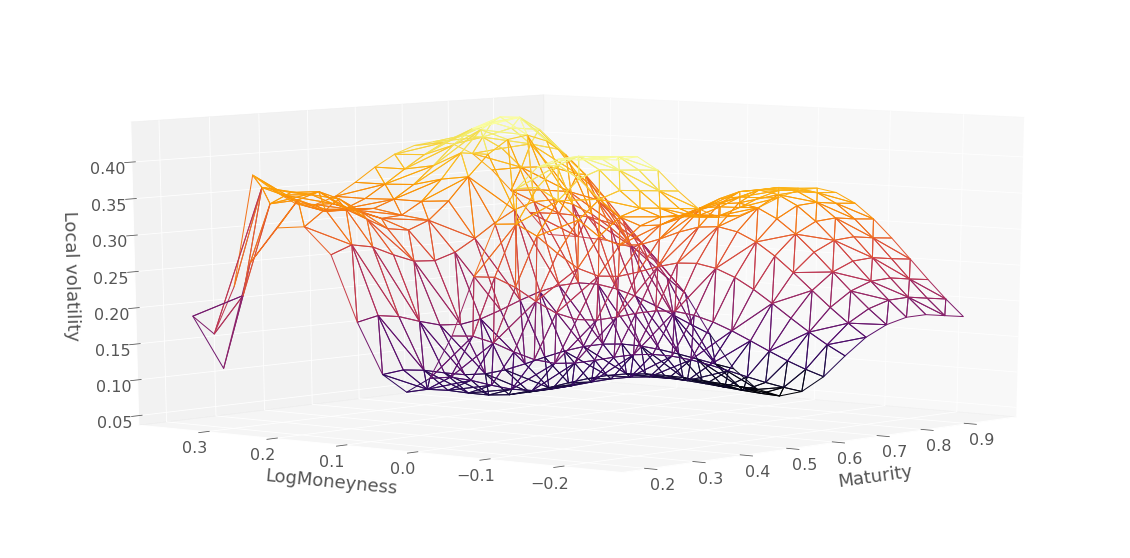}
  \caption{Hard Constraints}
  \label{fig:09082001HardReg}
\end{subfigure}
\begin{subfigure}{.48\textwidth}
  \centering
  \includegraphics[width=1.0\linewidth]{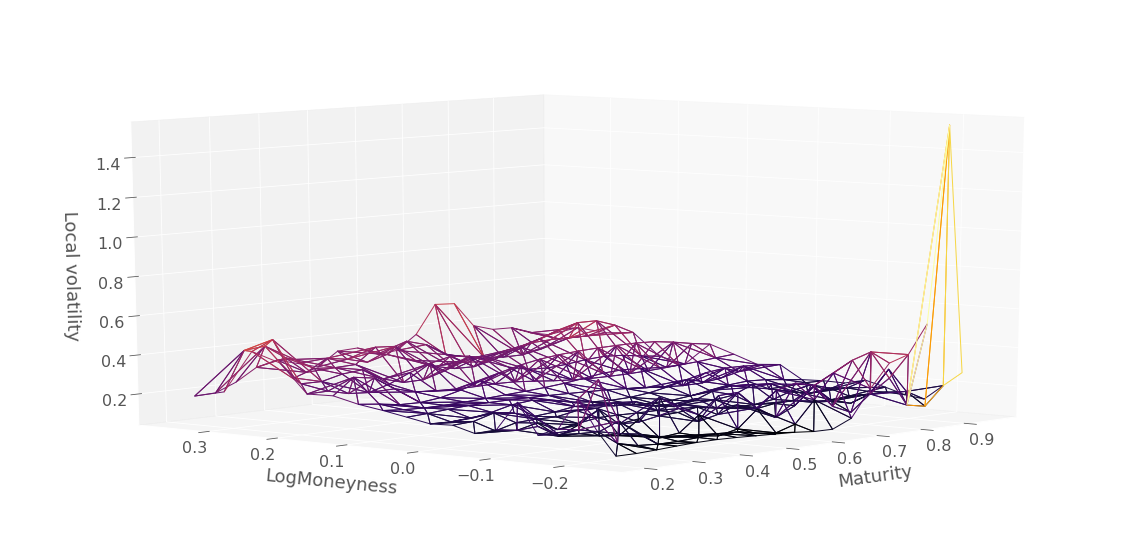}
  \caption{Soft constraints (w/o local vol. constraints)}
  \label{fig:09082001Soft2}
\end{subfigure}%

\begin{subfigure}{.48\textwidth}
  \centering
  \includegraphics[width=1.0\linewidth]{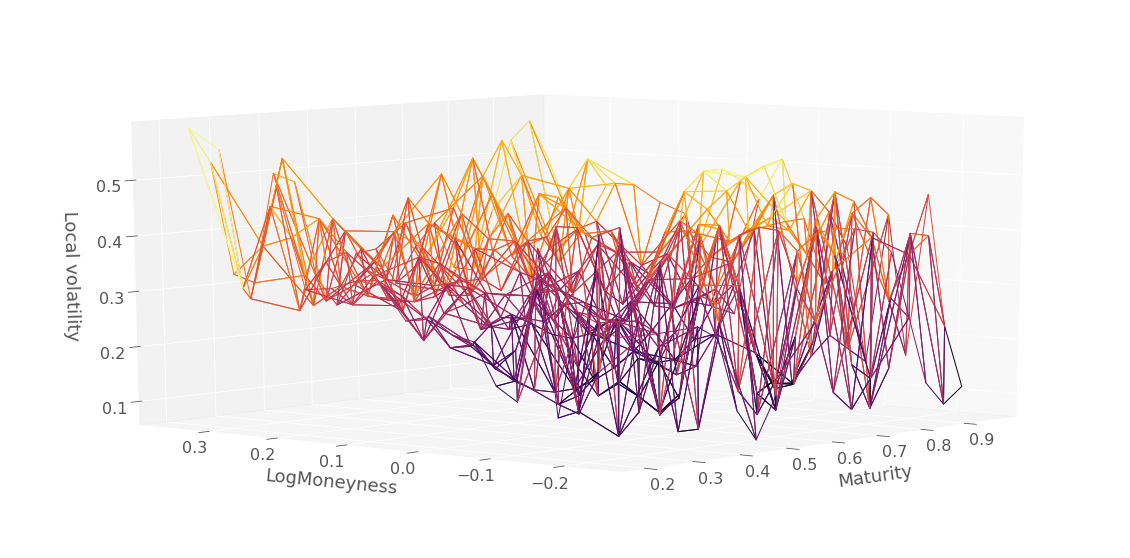}
  \caption{Soft constraints (with local vol. constraints)}
  \label{fig:09082001Soft2bis}
\end{subfigure}%
\begin{subfigure}{.48\textwidth}
  \centering
  \includegraphics[width=1.0\linewidth]{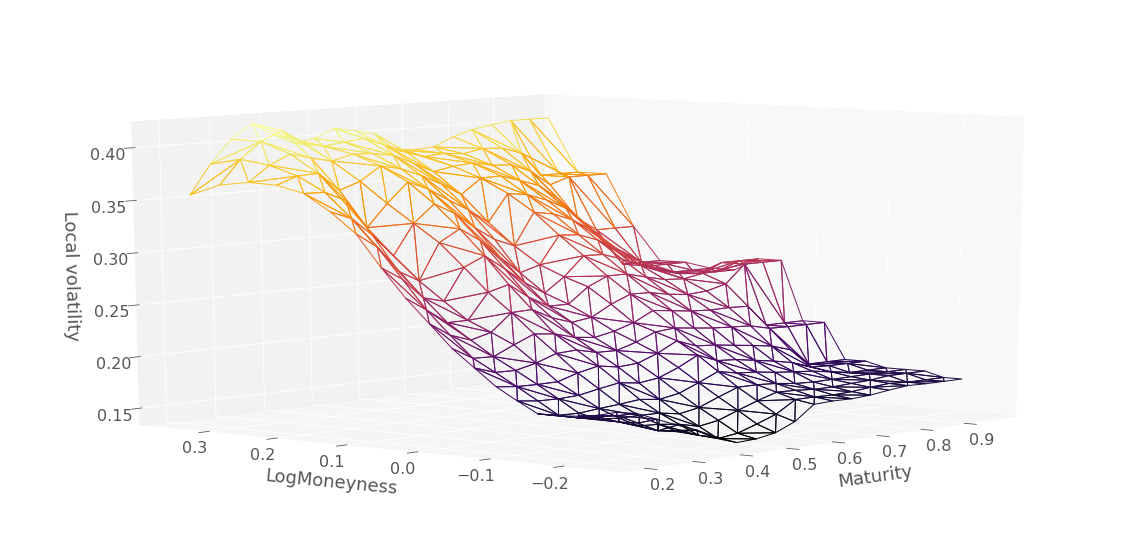}
  \caption{Tikhonov local volatility}
  \label{fig:09082001SoftTrue}
\end{subfigure}

\caption{\textit{Local volatility for 09/08/2001.}}
\label{fig:abc}
\end{sidewaysfigure}

Figures \ref{fig:testMulti12}, \ref{fig:testMulti13} and \ref{fig:abc} show the comparison of the local volatility surfaces obtained using hard constraints (sparse network) without Dupire penalization, dense network and soft constraints without and with Dupire penalization, as well as the Tikhonov regularization approach of \cite{crepey2002calibration}, on price quotes listed on August $7^{th}$, $8^{th}$, and $9^{th}$, 2001, respectively.
The soft constraint approach without Dupire penalization is both irregular (exhibiting 
outliers on a given day) and unstable (from day to day). In contrast, the soft constraint approach with Dupire penalization yields a more regular (at least, less spiky)
local volatility surface, both at fixed calendar time and in terms of stability across calendar time.
From this point of view the results are then qualitatively comparable to those obtained by Tikhonov regularization (which is however quicker, taking of the order of 30s to run).
 
\subsection{Monte Carlo Backtesting Repricing Error}

Finally, we evaluate the performance of the models in a backtesting Monte Carlo exercise. Namely, the options in each testing grid are repriced by Monte Carlo 
with $10^5$ paths of 100 time steps  in the model 
\be\label{e:lv} \frac{dS_t}{S_t} = \left( r(t) - q(t) \right) dt + \sigma(t,S_t) dW_t, \ee
using differently calibrated local volatility functions  
 $\sigma(\cdot,\cdot) $ in \eqref{e:lv}, for each of the 7th, 8th, and 9th August dataset.
 Table \ref{tab:MC} shows the corresponding Monte Carlo backtesting  repricing errors, using 
the option market prices from the training grids as reference values in the corresponding
RMSEs. The neural network approaches provide a full surface of prices and local volatilities, as opposed to values at the calibration trinomial tree nodes only in the case of Tikhonov, for which the Monte Carlo backtesting exercise thus requires an additional layer of local volatility inter-extrapolation, here achieved by a 
nearest neighbors algorithm.
We see from the table that both the benchmark Tikhonov method and the dense network soft constraints approach with Dupire penalization yield very reasonable and acceptable repricing errors (with still a certain advantage to the Tikhonov method), unlike the hard constraints approaches.
Moreover, the Dupire penalization is essential for extracting a decent local volatility function:
The dense network with soft constraint but without this penalization yields very poor Monte Carlo repricing RMSEs.

\begin{center}
\begin{table}[htbp!]
\resizebox{\columnwidth}{!}{%
\begin{tabular}{|c|c|c|c|c|c|}
\hline
$\sigma(\cdot,\cdot)$
 &\begin{tabular}[c]{@{}c@{}}Tikhonov \\ Monte Carlo\end{tabular}
   & \begin{tabular}[c]{@{}c@{}}Dense network \\ with soft constraints \\ and Dup. penal.\end{tabular} & \begin{tabular}[c]{@{}c@{}}Dense network \\ with soft constraints \end{tabular} & \begin{tabular}[c]{@{}c@{}}Hard constraint\\ with Dup. penal.\end{tabular} & \begin{tabular}[c]{@{}c@{}}Hard constraint\\ w/o Dup. Pen.\end{tabular} \\ \hline
07/08/2001                                                                 &  5.42 &   10.18  &    68.48                                                                  & 48.57                                                                 & 50.44   \\          \hline
08/08/2001   &                                                               5.55&    7.44   &   50.82                                                                   & 56.63                                                                 & 56.98    \\ \hline
09/08/2001  &                                                                4.60    & 8.18   & 59.39                                                                      & 66.23                                                                 & 65.50 \\ \hline
\end{tabular}}
\caption{Monte Carlo backtesting repricing RMSEs on training grid against market prices.}
\label{tab:MC}
\end{table}
\end{center}

The residual gap between the Monte Carlo RMSEs of the (even best) neural network local volatility and of the Tikhonov local volatility can seem disappointing.
However we should keep in mind that the neural network can evaluate quickly a local volatility on any node outside the training grid, whereas Tikhonov then requires a further layer of interpolation (or  a new calibration).
Furthermore, any vanilla option price can be accurately and quickly obtained by neural prediction (better than by Monte Carlo repricing as above).
Table \ref{tab:MCTrain}
shows training set
RMSEs of thus predicted prices against markets prices equivalent to (in fact, slightly better than) RMSEs
of Tikhonov trinomial tree prices against the same markets prices. These are of course only in-sample errors, but the additional findings of Table \ref{tab:MC} suggest that these good results are not just overfitting.
\begin{table}[htbp!]
\begin{center}  
\begin{tabular}{|c|c|c|}
\hline
$\sigma(\cdot,\cdot)$ 
&  Tikhonov  trin. tree  & \begin{tabular}[c]{@{}c@{}} NN pred. (dense network with \\ soft constraints  and Dup. penal). \end{tabular}    \\ \hline
07/08/2001                         & 2.42    & 2.66   \\          \hline
08/08/2001                                                              & 2.67   & 2.48  \\ \hline
09/08/2001                                                              & 2.45   & 2.34 \\ \hline
\end{tabular}
\end{center}
\caption{Training set RMSEs of Tikhonov trinomial tree vs. NN predicted prices against market prices.}
\label{tab:MCTrain}
\end{table}

\section{Conclusion} \label{sect:conclusion}

We introduced three variations of deep learning methodology to enforce 
no-arbitrage interpolation of European vanilla put option prices: (i) modification of the network architecture to embed shape conditions (hard constraints), (ii) use of shape penalization to favor these conditions (soft constraints), and (iii) additional use of local half-variance bounds in the penalization via the Dupire formula. 

Our experimental results 
confirm 
that hard constraints, although providing the only fail-safe approach to no-arbitrage approximation, reduce too much the representational power of the network numerically.
Soft constraints provide much more accurate prices and implied volatilities, while only leaving space for sporadic arbitrage opportunities, which are not only occasional but also very far from the money, hence do not necessarily correspond to monetizable arbitrage opportunities once liquidity is accounted for. Once the Dupire formula is included in the penalization, the corresponding local volatility surface is also reasonably regular, at fixed day, and stable, in terms of both out-of-sample performance at fixed day and dynamically from day to day. The performance of the neural network local volatility calibration method then gets close to the one of the classical Tikhonov regularization method of \cite{crepey2002calibration}, but not better. It is also slower. However,
the neural network provides the full surface of prices and local volatilities, as opposed to values at the nodes of a trinomial tree only under the approach of \cite{crepey2002calibration}.

We thus enrich the machine learning literature on neural networks metamodeling of vanilla option prices in three respects: first, by considering the associated local volatility, which is interesting both in itself and as a tool for improving the learning of the option prices in the first place; second, by working with real data; third, by systematically benchmarking our results with the help of a proven 
(both mathematically and numerically)  classical, non machine learning calibration procedure, i.e.
Tikhonov regularization.
In this article, we focused on machine learning schemes for extracting the local volatility from option {\em prices}. The use of  option {\em implied volatilities} will be considered in a further paper. 



\appendix
\section{Change of Variables in the Dupire Equation}\label{s:chvar}

 
Letting $g(T,K)  = P(T,K) \exp{\left( \int_{0}^{T} q dt\right)}$, note that  
 \begin{eqnarray*}
 &&\dK{P(T,K)} = \exp{\left( -  \int_{0}^{T} q dt\right)} \dK{g(T,K)} \sp 
  \gK{P(T,K)} = \exp{\left( -  \int_{0}^{T} q dt\right)} \gK{g(T,K)}\\&&
  \dT{P(T,K)}  = \exp{\left( -  \int_{0}^{T} q dt\right)} \left( \dT{g(T,K)} -  q g(T,K)\right).
 \end{eqnarray*}
The Dupire formula is then rewritten in terms of $g$ as
 \begin{equation*}
 \dT{g(T,K)} = \frac{1}{2} \sigma^2(T,K) K^2 \gK{g(T,K)} - (r-q) K \dK{g(T,K)}  .
\end{equation*} 
Through the additional change of variables
$\thep(T,k) = g(T,K)$, where
 $
 k =\exp{\left( -\int_{0}^{T} (r-q) dt\right)} K ,
$
  we obtain 
 \begin{eqnarray*}
 \dK{g(T,K)} &=& \dK{\thep(T,k)} = \dk{\thep(T,k)} \dK{k} = \exp{ \left( - \int_{0}^{T} (r-q) dt\right)} \dk{\thep(T,k)}\\ 
  \gK{g(T,K)} &=& \gK{\thep(T,k)} = \dk{\dK{\thep(T,k)}} \dK{k} = \exp{ \left( - 2 \int_{0}^{T} (r-q) dt\right)} \gk{\thep(T,k)}\\
  \dT{g(T,K)} &=&  \dT{\thep(T,k)} + \dT{k} \dk{\thep(T,k)} = \dT{\thep(T,k)} - (r-q) k \dk{\thep(T,k)}
 \end{eqnarray*}
and arrive at the modified Dupire equation: 
 \begin{equation*}
 \dT{\thep(T,k)} = \frac{1}{2} \sigma^2(T,K) k^2 \gk{\thep(T,k)}  \end{equation*}
conveniently written as the Dupire half-variance formula \eqref{eq:dupk}.

\section{Network Sparsity and Approximation Error Bound}\label{sect:approx}

We recall a result from \cite{Ohn_2019} which describes how the sparsity in a neural network affects its approximation error.

\begin{figure}[H]
\centering
 	\includegraphics[width=0.6\textwidth]{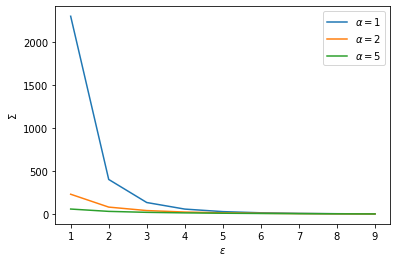}
	\caption{\textit{The upper bound on the network sparsity, $|\theta|_0 \leq \Sigma$, as a function of the error tolerance $\epsilon$ and H\"{o}lder smoothness, $\alpha$, of the function being approximated. As
	$\epsilon$ or $\alpha$ decrease, the value of $\Sigma$ is observed to increase. The plot is shown for $i=2$ and $\Sigma_0=10$.}}
\label{fig:holder}
\end{figure}

Let us denote the network parameters $\theta:=(\mathbf{W},\mathbf{b})$. We define the network parameter space in terms of the layers {width $L$ and depth $N$, and numbers of inputs and outputs,} $$\Theta_{i,o}(L,N):=\{\theta: L(\theta)\leq L, n_{max}(\theta) \leq N, in(\theta)=i, out(\theta)=o\}.$$ We also define the  restricted parameter space  
$$\Theta_{i,o}(L,N, \Sigma, B):=\{\theta \in \Theta_{i,o}  (L,N): |\theta|_0 \le  \Sigma, |\theta|_{\infty}
\le  B\},$$
where $|\theta|_0$ is the number of nonzero components in $\theta$ and $|\theta|_{\infty}$ is the largest absolute
value of elements of $\theta$.
Let the activation $\varsigma$ be either piecewise continuous or locally quadratic\footnote{A function $\varsigma:\mathbb{R}\rightarrow\mathbb{R}$ is locally quadratic if $\exists$ an open interval $(a, b) \subset \mathbb{R}$ over which $\varsigma$ is three times continuously differentiable with bounded derivatives and $\exists  t \in (a, b)$ s.t. $\varsigma'(t)\neq 0$ and $\varsigma''(t)\neq 0$.}. For example, softplus and sigmoid functions are locally quadratic. Let the function being approximated, $\thep\in \mathcal{H}^{\alpha,R}([0,1]^i)$ be H\"{o}lder smooth with parameters $\alpha>0$ and $R>0$, where $\mathcal{H}^{\alpha,R}(\Omega):=\{\thep \in \mathcal{H}^{\alpha}(\Omega): ||\thep||_{\mathcal{H}^{\alpha}(\Omega)} \leq R\}$. Then Theorem 4.1 in \cite{Ohn_2019} states the existence of positive constants $L_0, N_0, \Sigma_0, B_0$ depending only on $i, \alpha, R$ and $\varsigma$ s.t. for any $\epsilon>0$, the neural network
$$\theta_{\epsilon} \in \Theta_{i,1}\left(L_0log(1/\epsilon), N_0\epsilon^{-i/\alpha}, \Sigma_0\epsilon^{-i/\alpha}log(1/\epsilon),B_0\epsilon^{-4(i/\alpha+1)}\right)$$
satisfies $ \sup_{x\in[0,1]^i} | \thep(x)-\thep_{\theta_{\epsilon}}(x)|  \leq \epsilon$. Figure \ref{fig:holder} shows the upper bound $\Sigma$ on the network sparsity, $|\theta|_0\leq \Sigma$, as a function of the error tolerance $\epsilon$ and H\"{o}lder smoothness, $\alpha$, of the function being approximated. Keeping the number of neurons in each layer fixed, the graph shows that a denser network, with a higher upper bound, results in a lower approximation error. Conversely, networks with a low number of non-zero parameters, due to zero weight edges, exhibit larger approximation error. In the context of no-arbitrage pricing, the theorem suggests a tradeoff between using a sparse network to enforce the shape constraints, yet increasing the approximation error.  The adverse effect of using a sparse network should also diminish with increasing smoothness in the function being approximated.

\bibliographystyle{chicago}
\bibliography{main}
\end{document}

%% file: figures/nn2_many.tex
%


\def\layersep{1.5cm}

\begin{tikzpicture}[shorten >=1pt,->,draw=black!50, node distance=\layersep,transform shape,rotate=90]  
    \tikzstyle{every pin edge}=[<-,shorten <=1pt]
    \tikzstyle{neuron}=[circle,fill=black!25,minimum size=20pt,inner sep=0pt]
    \tikzstyle{input neuron}=[neuron, fill=yellow!50];
    \tikzstyle{output neuron}=[neuron, fill=red!50];
    \tikzstyle{hidden neuron}=[neuron, fill=green!50];
    \tikzstyle{annot} = [text width=4em, text centered]
    \tikzset{hoz/.style={rotate=-90}}   
      \node[input neuron, pin=left:\rotatebox{-90}{\parbox[t][][r]{8mm}{\centering  $T$}}] (I-1) at (0,-1) {};
      \node[input neuron, pin=left:\rotatebox{-90}{\parbox[t][][r]{8mm}{\centering  $k$}}] (I-2) at (0,-2) {};

    \foreach \name / \y in {1,...,10}
            \node[hidden neuron] (H1-\name) at (\layersep,-\y+4.0) {};



 \foreach \name / \y in {1,...,10}
            \node[hidden neuron] (H2-\name) at (2*\layersep,-\y+4.0) {};

      \foreach \name / \y in {1,...,1}
    \node[output neuron, pin={[pin edge={->}]right:\rotatebox{-90}{$\thep(T,k)$}}] (O-\name) at (3*\layersep,-\y-0.5) {};

      \foreach \dest in {1,...,5}
            \path (I-1) edge (H1-\dest);
      \foreach \dest in {6,...,10}
            \path (I-2) edge (H1-\dest);
            
    \foreach \source in {1,...,10}
        \foreach \dest in {1,...,10}
        \path (H1-\source) edge (H2-\dest); 
       \foreach \source in {1,...,10}
        \foreach \dest in {1,...,1}
            \path (H2-\source) edge (O-\dest);


\end{tikzpicture}